\documentclass[a4paper,fleqn,usenatbib]{mnras}

\usepackage[T1]{fontenc}
\usepackage{ae,aecompl}

\usepackage{graphicx}   
\usepackage{amsmath}    
\usepackage{amssymb}    

\title[Radio and optical intra-day variability observations of five blazars]{
Radio and optical intra-day variability observations of five blazars}
\author[X.~Liu et al.]{X.~Liu\(^{1,2}\)\thanks{E-mail:
liux@xao.ac.cn}, P.P.~Yang\(^{1.3}\), J.~Liu\(^{1,2}\), B.R.~Liu\(^{4}\), S.M.~Hu\(^{5}\), O.M.~Kurtanidze\(^{6,7,8}\), S.~Zola\(^{9,10}\),
\newauthor A.~Kraus\(^{11}\), T.P.~Krichbaum\(^{11}\), R.Z.~Su\(^{1.3}\), K.~Gazeas\(^{12}\), K.~Sadakane\(^{13}\), K.~Nilson\(^{14}\),
\newauthor D.E.~Reichart\(^{15}\), M.~Kidger\(^{16}\), K.~Matsumoto\(^{13}\), S.~Okano\(^{17}\),
\newauthor M.~Siwak\(^{10}\), J.R.~Webb\(^{18}\), T.~Pursimo\(^{19}\), F.~Garcia\(^{20}\), R. Naves~Nogues\(^{21}\),
\newauthor A.~Erdem\(^{22,23}\), F.~Alicavus\(^{22,23}\), T.~Balonek\(^{24}\), S.G.~Jorstad\(^{25}\)\\
\(^{1}\) Xinjiang Astronomical Observatory, Chinese Academy of Sciences, 150 Science 1-Street, Urumqi 830011, China\\
\(^{2}\) Key Laboratory of Radio Astronomy, Chinese Academy of Sciences, Nanjing 210008, China\\
\(^{3}\) Graduate University of Chinese Academy of Sciences, Beijing 100049, China\\
\(^{4}\) Guangxi Key Laboratory for Relativistic Astrophysics, Guangxi University, Nanning, Guangxi  530004, China\\
\(^{5}\) Shandong Provincial Key Laboratory of Optical Astronomy and Solar-Terrestrial Environment, Institute of Space Sciences, Shandong University,\\ Weihai, 264209 Weihai, China\\
\(^{6}\) Abastumani Observatory, Mt. Kanobili, 0301 Abastumani, Georgia\\
\(^{7}\) Engelhardt Astronomical Observatory, Kazan Federal University, Tatarstan, Russia\\
\(^{8}\) Landessternwarte Heidelberg, K\"onigstuhl 12, 69117 Heidelberg, Germany\\
\(^{9}\) Astronomical Observatory, Jagiellonian University, ul. Orla 171, PL-30-244 Krakow, Poland\\
\(^{10}\) Mt. Suhora Observatory, Pedagogical University, ul. Podchorazych 2, PL-30-084 Krakow, Poland\\
\(^{11}\) Max-Plank-Institut f\"ur Radioastronomie, Auf dem H\"ugel 69, 53121, Bonn, Germany\\
\(^{12}\) Section of Astrophysics, Astronomy and Mechanics, National \& Kapodistrian University of Athens, Zografos GR-15784, Athens, Greece \\
\(^{13}\) Astronomical Institute, Osaka Kyoiku University, 4-698 Asahigaoka,Kashiwara, Osaka 582-8582, Japan\\
\(^{14}\) Tuorla Observatory, Department of Physics and Astronomy, University of Turku, Turku F-21500, Finland\\
\(^{15}\) Department of Physics and Astronomy, University of North Carolina, Chapel Hill, North Carolina 27599, USA \\
\(^{16}\) Herschel Science Centre, ESAC, European Space Agency, C/Bajo el Castillo, s/n, Villanueva de la Can\(\tilde{a}\)da, Madrid E-28692, Spain \\
\(^{17}\) Planetary Plasma and Atmospheric Research Center, Tohoku University, Sendai 980-8578, Japan \\
\(^{18}\) Florida International University and SARA Observatory, University Park Campus, Miami, Florida 33199, USA\\
\(^{19}\) Nordic Optical Telescope, Apartado 474, Santa Cruz de La Palma E-38700, Spain \\
\(^{20}\) Mun\(\tilde{a}\)s de Arriba La Vara, Vald\(\acute{e}\)s  E-33780, Spain \\
\(^{21}\) C/Jaume Balmes No 24, Cabrils, Barcelona E-08348, Spain \\
\(^{22}\) Department of Physics, Faculty of Arts and Sciences, Canakkale Onsekiz Mart University, Canakkale TR-17100, Turkey \\
\(^{23}\) Astrophysics Research Center and Ulupinar Observatory, Canakkale Onsekiz Mart University, Canakkale TR-17100, Turkey \\
\(^{24}\) Foggy Bottom Observatory, Colgate University, 13 Oak Drive, Hamilton, New York 13346, USA \\
\(^{25}\) Institute for Astrophysical Research, Boston University, 725 Commonwealth Ave., Boston, Massachusetts 02215, USA\\
}

\date{Accepted XXX. Received YYY; in original form ZZZ}

\pubyear{2017}

\makeatletter
\AtBeginDocument{\let\($\let\)$}
\makeatother
\begin{document}

\label{firstpage}
\pagerange{\pageref{firstpage}--\pageref{lastpage}}
\maketitle

\begin{abstract}
We carried out a pilot campaign of radio and optical band intra-day variability (IDV) observations of five blazars (3C66A, S5 0716+714, OJ287, B0925+504, and BL Lacertae) on December 18--21, 2015 by using the radio telescope in Effelsberg (Germany) and several optical telescopes in Asia, Europe, and America. After calibration, the light curves from both 5 GHz radio band and the optical R band were obtained, although the data were not smoothly sampled over the sampling period of about four days. We tentatively analyse the amplitudes and time scales of the variabilities, and any possible periodicity. The blazars vary significantly in the radio (except 3C66A and BL Lacertae with only marginal variations) and optical bands on intra- and inter-day time scales, and the source B0925+504 exhibits a strong quasi-periodic radio variability. No significant correlation between the radio- and optical-band variability appears in the five sources, which we attribute to the radio IDV being dominated by interstellar scintillation whereas the optical variability comes from the source itself. However, the radio- and optical-band variations appear to be weakly correlated in some sources and should be investigated based on well-sampled data from future observations.
\end{abstract}




\section{Introduction}

Among active galactic nuclei (AGNs), blazars (a combination of flat-spectrum radio quasars and BL Lac objects) exhibit extreme variations in flux over almost the entire electromagnetic spectrum on time scales ranging from less than hours to many years. This violent behavior of blazars is generally explained in terms of the shock-in-jet model \citep{mars85} and the beaming effect, in which a relativistic jet is oriented very close to our line of sight \citep{urry95}.

Discovered in the 1980s \citep{witz86,hees87}, the intra-day variability (IDV) of AGNs at centimeter wavelengths occurs in 56\% of the flat-spectrum radio AGNs \citep{love08} and in \(\sim\)60\% of the \(\gamma\)-ray blazars \citep{liu12a}. The cm-radio IDV has been largely attributed to the interstellar scintillation (ISS) for the dependence of variability on the Galactic latitude and the H$_{\alpha}$ emission (the Wisconsin H Alpha Mapper (WHAM) Northern sky survey, \cite{haff03}), see \cite{love08}, and especially for the annual modulation (owing to the Earth revolution around the Sun) of variability time scales discovered in a couple of IDV sources \citep[e.g.][]{jaun01,rick01,denn03,big03,big06,gaban07,liu12b,liu13} and the time delays of IDV found in trans-continent observations \citep{jaun00,denn02,big06}. For more details, see the review by \cite{jaun16}.

In the optical band, however, IDV should be intrinsic to the source because ISS does not affect extragalactic emission in optical wavelengths.

It is generally believed that the brightness temperature \(T_{\rm b}\) of the radio-jet component of AGNs cannot exceed the inverse Compton limit \(T_{\rm IC}=10^{11.5}\) K \citep{kell69} or the equipartition temperature \(T_{\rm eq}=5\times 10^{10}\simeq10^{10.7}\) K \citep{read94}. If radio IDV at cm wavelengths is source-intrinsic, the brightness temperature of emitting components will exceed the limit \(T_{\rm IC}\) or \(T_{\rm eq}\) by several orders of magnitudes, which would require a very large Doppler factor \(\delta\) (e.g., \({>} 100\)). However, for blazars, Doppler factors less than 30 are usually estimated from Very Long Baseline Interferometry (VLBI) observations or flux monitoring data \citep{liod17}.

Although cm-radio IDV is generally attributed to ISS, source-intrinsic radio IDV might still be possible to detect in the following situations: (1) In some blazar cores, the intrinsic brightness temperature \(T_{\rm int}=T_{\rm obs}/\delta\) is greater than \(10^{12}\) K (where \(T_{\rm obs}\) is measured from the RadioAstron space VLBI observations and \(\delta\) is estimated from ground VLBI arrays; see \cite{kova16}), implying that \(T_{\rm IC}\) could be violated for these blazars \citep{mars16}, but such apparent detections of high temperatures on very long baselines as seen by RadioAstron may also be caused by refractive substructure
introduced by scattering in the interstellar medium, which is more evident at 1.6 GHz \citep{john16}. (2) A large fraction of IDV sources are intermittent \citep{ked06,love08}, which should allow source-intrinsic IDV to be observed when ISS disappears. (3) Sometimes the radio and optical IDVs seem to be correlated, e.g., in blazar S5 \(0716+714\), \cite{quir91} found, during a four-week monitoring campaign in 1990, simultaneous transitions from fast to slow variability modes among 5 GHz radio-band and optical-band wavelengths, they interpret the similar transitions as source intrinsic variability.

However, the existence of source-intrinsic IDV, if present, is difficult to disentangle from ISS-induced variability, and it may require multi-wavelength measurements with, for example, radio, optical, X-ray, and \(\gamma\)-ray IDV observations to search for correlations between the bands. Note that the optical and high-energy-band IDVs are thought to be intrinsic to the source. Very few multi-band IDV observations have been made over the past decades. To address this shortcoming, we are planning more radio-optical IDV observations. As a first step in this plan, we carried out a pilot campaign on December 18--21, 2015, using two radio telescopes and several optical telescopes.


\section[]{Observations and data reduction}

We carried out quasi-simultaneous radio-optical observations of
five blazars on December 18--21, 2015 using the following
 telescopes: Effelsberg and Urumqi (5 GHz radio),
Weihai (1 m), Urumqi (1 m), Abastumani (70 cm), the Blazar Optical Sky Survey
(BOSS Project) at the University of Athens Observatory (40 cm),
and OJ287-15/16 Collaboration telescopes (optical R band). The target sources were 3C66A, S5 \(0716+714\), OJ287, B\(0925+504\), and BL Lac, which were selected from the RadioAstron 2015 December observing sample, except B\(0925+504\) that we added in this campaign for its significant radio IDV observed before. These sources are also bright in optical, so we want to compare their radio and optical variations, and it will be interesting to see if there is a radio-optical correlation for S5 \(0716+714\) as suggested before \citep{quir91}. Unfortunately, the Urumqi radio telescope was being reconstructed and was still in testing at that time. It suffered from serious problems involving pointing, etc., that prevented the data from being used in our analysis. In addition, the Urumqi optical observation had a problem with its camera gate, which led to serious light pollution that prevented that data from being used as well.

The radio observations at Effelsberg, however, were successful. These observations were performed by cross-scans in azimuth and elevation (all sources were point-like for the 100-m telescope). For these observations, the source was actually scanned two times in both azimuth and elevation. After initial calibration of the raw data, the
intensity profile of each scan was fit with a Gaussian
function after subtracting the baseline. The fitted scans were then
averaged over both azimuth and elevation. Next, after pointing correction, the azimuth and elevation
scans were averaged together, and then corrections for the gain-elevation effect of the antenna
and the (rather small) atmospheric absorption were applied, with the secondary calibrators monitored during
the observation. These calibrators were also used to correct
the data for systematic time-dependent effects, with an accuracy of \(\sim\)0.4\% (1\(\sigma\)) of total flux density. Finally, the
amplitudes were converted to absolute flux density by using
the average scale of the primary calibrators (3C48, 3C286, and NGC7027). The observation lasted nearly four days, with a typical sampling rate of 1--2 hours for
each calibrated data point. However, because sources were not always in the sky for 24 hours, some gaps appear in the light curves of the sources.

The raw optical data were processed by using standard methods (e.g., by using the standard routines of the Image Reduction and Analysis Facility or DAOPHOT II calibrated with comparison stars in the observation field) (see \cite{valt16} for the data-reduction method of the OJ287-15/16 Collaboration). We do not describe in detail here the telescopes and cameras involved in these observations. However, note that, because the sources are
compact, the different apertures used at different telescopes (\(\leq\)1 m) had
a negligible effect on the measured flux. Photometric error was estimated from the root-mean-square (rms) deviations of the reference stars and found to be generally less than 0.02 mag, although some larger errors appeared in some time slots of individual telescopes (e.g., in the last day of the Weihai observations), which were mostly due to bad weather.

\section[]{Results and variability characteristics}

\subsection[]{Light curves and variability analysis}

Figures \ref{fig1}--\ref{fig4} and \ref{fig6} show the radio and optical light curves, for 3C66A, S5 \(0716+714\), OJ287, B\(0925+504\), and BL Lac, respectively. We used the following methods to analyse the variability:
\begin{itemize}
\item[(1)] The relative amplitude of variability, which is defined as the maximum \(S_{\rm max}\) minus the minimum \(S_{\rm min}\) divided by the maximum plus the minimum of a light curve \citep{kova05}:
\begin{equation}
V = (S_{\rm max}-S_{\rm min})/(S_{\rm max}+S_{\rm min}).
\end{equation}

The results are shown in  Table~\ref{tab1} for both the radio and optical variability for the five sources. We also used the rms flux variation (in percent), which is the ratio of the rms variability \(\sigma_{\rm var}\) divided by the mean flux density \(\overline{S}\): \(m=\sigma_{\rm var}/\overline{S}\). Note that the units of the optical magnitude are not those of the power intensity of the source, such as the radio-flux density; here we are concerned only with the relative variations and do not intend to transform these variations into flux density. Note that, in Table \ref{tab1}, \(m<V\) by definition for both radio and optical variations.

\item[(2)] We analyse the variability time scale \(t\) and possible variability period \(P\) in the light curves. For this, we used the first-order standard
structure function (SF); see \cite{sim85}. The SF usually has a power-law form that
reaches its maximum at a saturation level. Here we define the intersection of
the power-law fit with the plateau corresponding to this saturation
level as the time scale of the characteristic variability. The error
of the time scale can be estimated by accounting for
the formal errors of the power-law fit and the fitting to the
SF plateau. Sometimes the SF has more
than one plateau, in which case multiple variability
time scales are estimated.

To determine the possible periodicity of the variations in light curves, we applied the power spectrum density (PSD) analysis, in which strong narrow peaks in the PSD are assumed to be suggestive of variability periods, and the error is estimated from the full width at half power of the peak. Only three sources (S5 \(0716+714\), OJ287, B\(0925+504\)) show possible periodic variations (see Table~\ref{tab1}).

Note that the resulting variability time scales and periods are tentative because of the gaps in the light curves of the sources (e.g., 3C66A, OJ287 and BL Lac in radio band, and 3C66A, S5 \(0716+714\), B\(0925+504\) and BL Lac in optical band).

\item[(3)] The correlation between radio and optical light curves was analysed by using the Pearson correlation coefficient and a null probability or confidence level \citep[e.g.][]{olk58,fan17}; the result is listed in Table~\ref{tab1}. The results reveal no significant correlations between the radio and optical variations for the five blazars, although weak correlation seems to appear in some cases (S5 \(0716+714\), B\(0925+504\)), but with low significance.
\end{itemize}

\begin{table*}

 \caption[]{Variability amplitudes and possible variability time scales and periods of blazars, and the correlation parameter between radio and optical variations. The columns are (1) source name, (2) redshift, (3) observation band (5 GHz radio in first line, optical R band in second line), (4) rms amplitude variation divided by average value in percent, (5) relative amplitude variations in percent as defined in the text, (6) variability time scales (in days) estimated from the SF, (7) variability periods (in days) estimated from the power spectrum density, (8) Pearson correlation coefficient \(r\), (9) null probability \(p_{null}\). The calibrator's modulation index \(m_{0}\) is 0.35\% at radio band.}
         $$
         \begin{tabular}{c|c|c|c|c|c|c|c|c}
\hline
  \hline
    (1)&(2) &(3) &(4) & (5)&(6) &(7) & (8) & (9)\\
 \(Source\) & z & \(Band\) & \(m\) (\%) & \(V\) (\%) & \(t(d)\) &  \(P(d)\)  & \(r\) & \(p_{null}\) \\
    & &  & & &  &
    &  & \\

\hline

  3C66A  &0.444 & Radio  & 1.5 & 4.7  & \({>}3.0\)  &            &   0.07                  &  0.81         \\

         & & Optical   & 0.3 & 1.2   & 0.6(0.1), 1.4(0.2)  &           &      &            \\

           \hline

 S5 \(0716+714\) &\(\sim\)0.3 & Radio  & 3.6 & 7.4 &  2.3(0.2)       &        &    0.35                    &   0.13       \\
          & & Optical  & 0.6 & 1.5 & 0.5(0.1)        &   0.8(0.2)      &        &           \\

            \hline

 OJ287   & 0.306& Radio   & 3.0 & 7.3 &  0.4(0.1), \({>}1.8\)       &       &   \(-0.47\)                       &  0.04    \\

         & & Optical   & 1.2 & 3.4 & 0.3(0.1), 1.0(0.1)  &   0.5(0.1), 1.3(0.2)    &         &           \\

           \hline

 B\(0925+504\)& 0.37& Radio  & 5.2 & 10.0 & 0.3(0.1)     &    0.5(0.1)         &      0.34      &    0.26    \\

         & & Optical   & 1.1 & 2.5 &   0.3(0.1)      &            &         &          \\

           \hline

 BL Lac  &0.0686 & Radio     & 1.1 &  2.4     &   1.9(0.2)   &     &    0.32                      &  0.21           \\

         & & Optical       &  0.6 & 1.6     &   0.4(0.1), 1.1(0.1)    &      &             &     \\

            \hline\hline
           \end{tabular}{}
         $$
         \label{tab1}
   \end{table*}

\section{Comments on individual sources}

All five blazars are \(\gamma\)-ray loud BL Lac objects and three of them are also TeV emitters (3C66A, S5 \(0716+714\), and BL Lac), and their radio spectra are flat at centimeter wavelengths. Four of them show kpc-scale core-halo features but not more than 100 kpc, and one (B\(0925+504\)) is compact and unresolved with the Very Large Array (VLA). The BL Lac objects are low-accreting AGNs and could be in the early evolutionary stage of becoming larger-scale Fanaroff--Riley Class I radio sources. In the following, we briefly discuss each the five sources based on the literature and on the present results.\\

3C66A, 2FGL J\(0222.6+4302\) (ISP BL)\\

This is an intermediate synchrotron peaked BL Lac
object \citep[ISP BL,][]{abd10,fan16} and a TeV \(\gamma\)-ray source. Optical variability at short time scales has been reported for this source \citep{xie94}. In the radio band, it appears as a core halo on the \(<\)100 kpc scale and has a one-sided VLBI core jet to the south, in which fast
apparent superluminal motions of up to 29\(c\) were reported by \cite{jors01}. No evident correlation between long-term \(\gamma\)-ray by the Fermi Large-Area Telescope and multiband radio emission was found in 3C66A\citep{fuhr14}.

\begin{figure}
  \includegraphics[width=8cm]{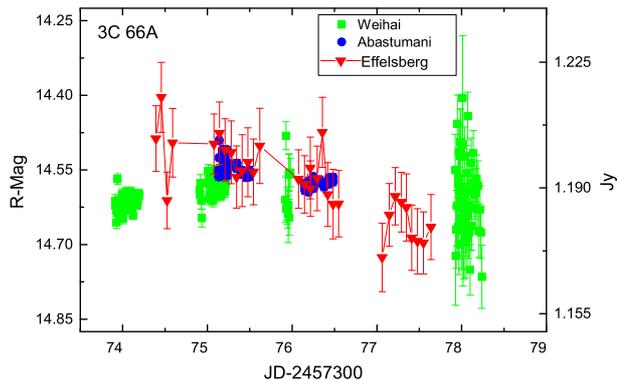}
  \caption{Radio-band (5 GHz) light curve (Effelsberg, on right axis) and optical R-band light curve (Weihai, Abastumani, on left axis) versus JD-2457300, for 3C66A.}
  \label{fig1}
\end{figure}

The present results show a marginal inter-day variability at 5 GHz (with reduced Chi-squared of 1.2) and rapid optical variations with time scales of \(0.6\pm0.1\) and \(1.4\pm0.2\) days (see Fig.~\ref{fig1} and Table~\ref{tab1}), and no correlation between the radio and optical variations is found.\\

S5 0716+714, 2FGL J\(0721.9+7120\) (ISP BL)\\

This object is a TeV \(\gamma\)-ray blazar and displays optical short-term variability and micro variability \citep{hu14,man16}. In the radio band, it shows a two-sided halo on the kpc scale, and a one-sided VLBI core jet to the north with highly relativistic speeds \citep{rani15,lis16}. A significant positive correlation exists between the position angle and the peak flux density of the inner jet, which is suggestive of a helical jet \citep{liu12c}. The VLBI jet-position angle and the \(\gamma\) rays are also correlated, which suggests that  radio waves and \(\gamma\) rays are emitted on the same helical filament, which rotates with the jet \citep{rani14}, \cite{rani15} further state that ``our analysis favours the major optical/\(\gamma\)-ray flares in S5 \(0716+714\) being produced upstream of the 7 mm VLBI core''. This statement makes it clear that the radio and optical/\(\gamma\)-ray flares in S5 \(0716+714\), come from separate regions. During a four-week monitoring campaign in 1990, simultaneous transitions were observed from fast to slow variability modes among 5 GHz radio and optical bands, suggesting a source intrinsic origin of the inter-day variability \citep{quir91}. \cite{fuhr08} found that the increase of the variability amplitudes from cm to mm bands contradicts expectations from standard ISS and suggests a source-intrinsic origin for the inter-day variability. \cite{gupta12} observed multi-waveband IDV and found no significant correlation between cm-radio and optical-band IDV. At 5 GHz, \cite{liu12b} found strong evidence in the form of an annual cycle of IDV time scales of S5 \(0716+714\), which suggests that the 6 cm IDV is dominated by ISS.

\begin{figure}
  \includegraphics[width=8cm]{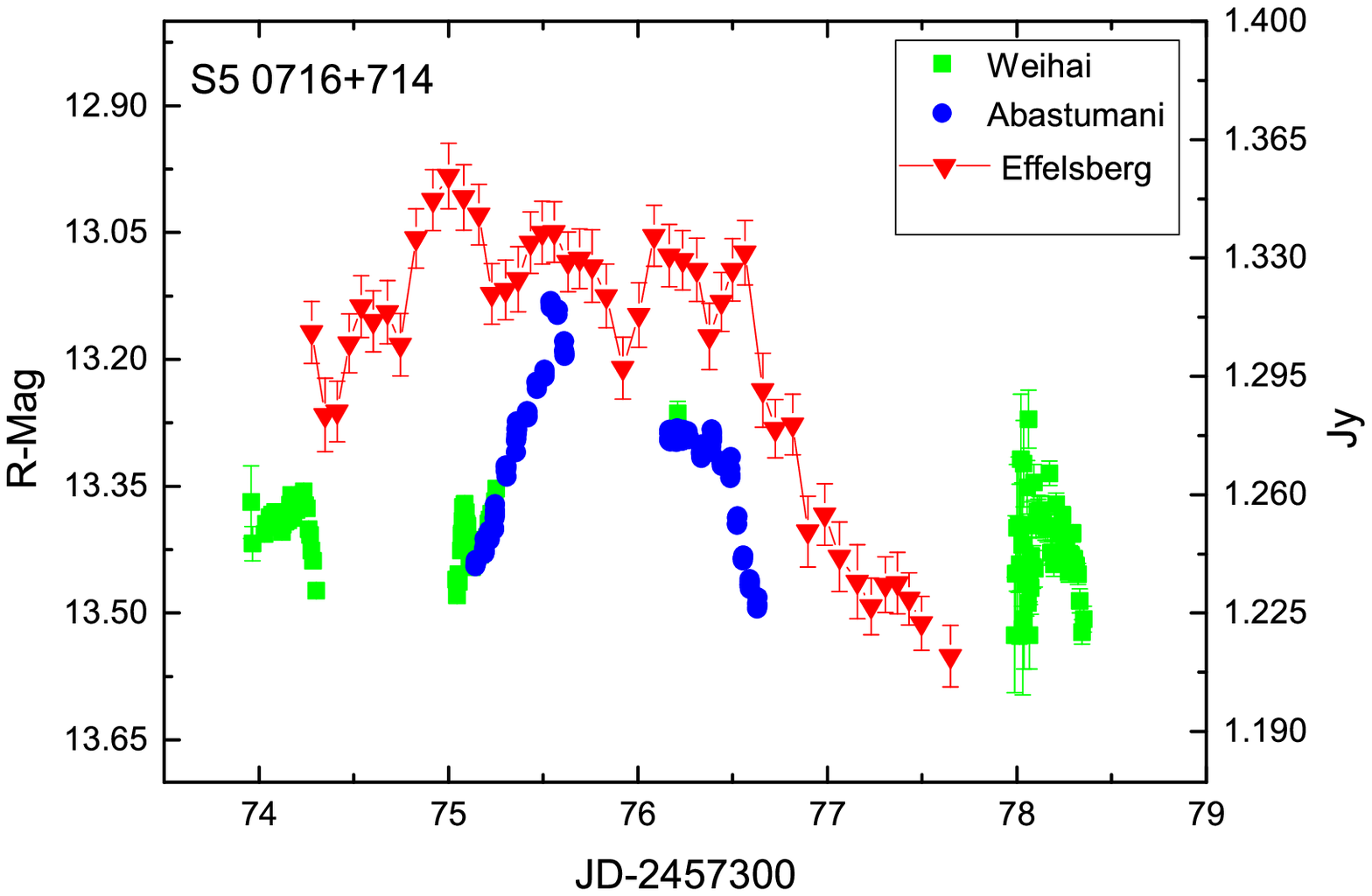}
  \caption{Radio-band (5 GHz) light curve (Effelsberg, on right axis) and optical R-band light curve (Weihai, Abastumani, on left axis) versus JD-2457300, for S5 0716+714.}
  \label{fig2}
\end{figure}

The present results show strong IDVs in both the radio and optical bands (Fig.~\ref{fig2} and Table~\ref{tab1}). The radio flux initially increases and then decreases, and contains small IDV wiggles. In the optical band, it undergoes pronounced variations. A weak radio-optical correlation exists in the `common radio-optical data time', but this is not significant (Table~\ref{tab1}).\\

OJ287, 2FGL J\(0854.8+2005\) (LSP BL)\\

The BL Lac object is known for its \(\sim\)12-year
periodicity in the optical light curve \citep{silla88}. The optical
outbursts appear as double peaked and their periodic occurrence is
interpreted primarily in terms of a binary black hole system \citep{valt07}. The model predicted a major optical outburst in  December, 2015, which occurred within the expected time range, peaking on December 5, 2015 at a magnitude of 12.9
in the optical R band (Valtonen et al. 2016). Various time scales and periods (or quasi-periodic oscillations) have been reported in the optical band \citep{sand16,bhat16,zola16}. In the radio band, it shows a one-sided core jet on the kpc scale to the west, similar to its VLBI jet direction \citep{per94}. A prominent erratic wobbling of the inner-jet-position angle changes by up to \(40^{\circ}\) over two years \citep{agu12}, and it has been suggested that OJ287 is a rotating helix \citep{cohen17}.
.

\begin{figure}
  \includegraphics[width=8cm]{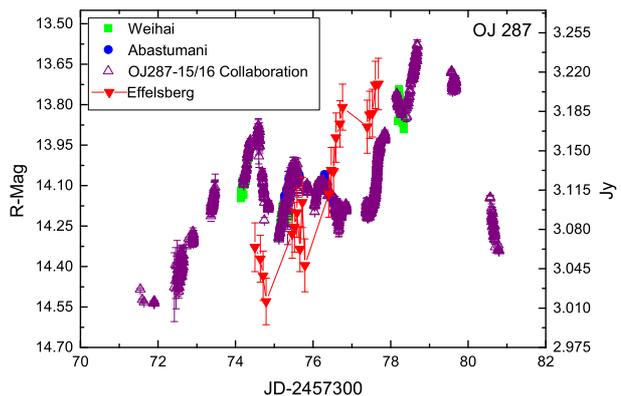}
  \caption{Radio (5 GHz) light curve (Effelsberg, on right axis) and optical R-band light curve (Weihai, Abastumani, OJ287-15/16 Collaboration, on left axis) versus JD-2457300, for OJ287.}
  \label{fig3}
\end{figure}

The present results show prominent intra- and inter-day variations in the radio and optical bands, with the shortest time scale being a few hours in the optical band (Fig.~\ref{fig3} and Table~\ref{tab1}). The radio-band light curve has three gaps, so the radio-band time-scale assignment is only tentative. Finally, a weak (non-significant) anti-correlation appears between the radio and optical variations in the common radio-optical data time (see Table~\ref{tab1}).\\

B\(0925+504\), 2FGL J\(0929.5+5009\) (ISP BL)\\

 The optical R-band image of this ISP BL Lac object at redshift 0.37 shows a very compact host galaxy that can be fitted with the de Vaucouleurs model \citep{nils03}. In the radio band, it is not resolved at 5 GHz with the VLA \citep{taylor96} but is resolved with the VLBI, showing a core jet to the southeast. Rapid radio IDVs have previously been detected towards this source at 5 GHz \citep{koay11,liu15}.

\begin{figure}
  \includegraphics[width=8cm]{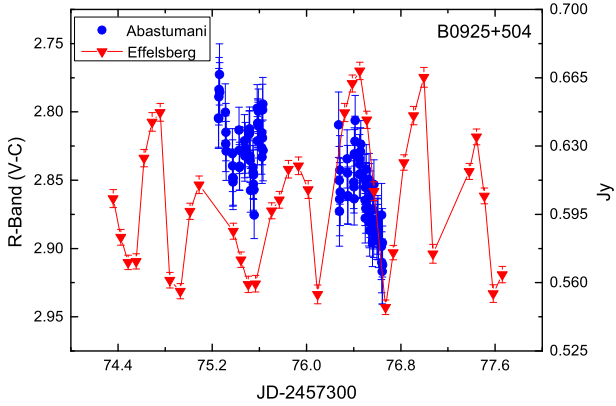}
  \caption{Radio (5 GHz) light curve (Effelsberg, on right axis) and optical R-band (V-C) light curve (Abastumani, on left axis) versus JD-2457300, for B\(0925+504\).}
  \label{fig4}
\end{figure}

\begin{figure}
  \includegraphics[width=8cm]{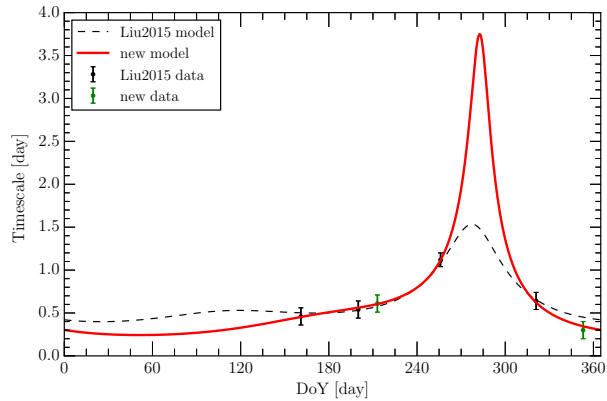}
  \caption{Radio (5 GHz) IDV timescales from Liu \& Liu (2015) and new observations (green) that we made in July and December 2015 (the campaign reported herein) against day of year for B\(0925+504\). The red solid line represents the updated model-expected timescales. See text for the values of the ISS model parameters.}
  \label{fig5}
\end{figure}

Our results show prominent intra- and inter-day variations with similar time scales of \(0.3\pm0.1\) day in the radio and optical bands (but the optical time scale is tentative because the data are from only a single site), and exhibit a remarkable quasi-periodic radio variability with \(P\sim0.5\pm0.1\) day estimated from the light curve (see Fig.~\ref{fig4} and Table~\ref{tab1}). \cite{liu15} find that, from our previous observations at Urumqi and Effelsberg, the IDV time scales of this source can be fitted with an ISS model, with the time scales mostly \(<\)0.5 day and the longer time scales becoming relevant only during days 250--300 of the year. The IDV time scale of \(0.3\pm0.1\) days observed on December 18--21, 2015 about day 355 of the year, is shorter than that predicted by the model. We combined all previous and current observations of this source and fit the time scales again with the model of the anisotropic scattering `screen' (see Fig.~\ref{fig5}). The parameters of the ISM model are updated with the projected velocity of the screen, \(V_{\rm RA}=-4.2\pm 1.1\) km/s, \(V_{\rm Dec}=-2.5\pm 1.8\) km/s, the ratio of the two-dimensional screen axis \(R=1.5\pm 0.3\), and the position angle of the major axis of the screen (measured from the north to east), \(\psi=111.1\pm 8.4\degr\). Note that the data are fit with \(D=110\) pc, with the assumption that the distance \(D\) from the ISM to Earth did not vary too much over the past years. See \cite{march12} and \cite{liu15} for more details of fitting to the ISS model.

Furthermore, a weak but non-significant radio-optical correlation exists (see Table~\ref{tab1}) for the common radio-optical data time (note, however, that only two nights of optical data are available). Further study of this correlation requires a longer optical-data train. However, the fitted annual cycle in Fig.~\ref{fig5} is clear evidence of ISS as a major cause of the radio IDV on B\(0925+504\).\\

BL Lacertae, 2FGL J\(2202.8+4216\) (LSP BL)\\

BL Lacertae (the prototype of BL Lac objects) lies at the centre of a giant elliptical galaxy with weak emission lines and is a TeV \(\gamma\)-ray source. The source exhibits optical variabilities with multiple time scales \citep{fan99}. In the radio band, it is a kpc-scale core halo and has a VLBI core jet to the south. The radio emission varies at higher frequencies, leading the lower-frequency emissions by several days to a few months \citep{villa04}, and the optical variations lead the 15 GHz radio emissions by \(\sim\)100 days \citep{bach06}.

\begin{figure}
  \includegraphics[width=8cm]{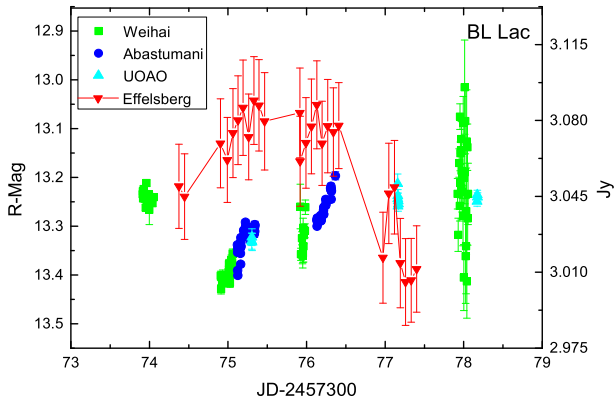}
  \caption{Radio-band (5 GHz) light curve (Effelsberg, on right axis) and optical R-band light curve (Weihai, Abastumani, the University of Athens
Observatory, on left axis) versus JD-2457300, for BL Lac.}
  \label{fig6}
\end{figure}

The present results (Fig. \ref{fig6}) show some intra- and inter-day optical variations with relatively weak variations in the radio band (with reduced Chi-squared of 1.5 and \(m=1.1\)\% in Table~\ref{tab1}, about \(3\sigma\) detection). A weak but non-significant correlation exists between radio- and optical-band variations in the common radio-optical data time (Table~\ref{tab1}).

\section{Discussion and summary}

Striking highly periodic radio-band variations of B\(0925+504\) are detected in this campaign, which share with the optical band a similar variability time scale of \(\sim\)0.3 days to the optical one. Furthermore, a weak positive correlation between the radio- and optical-band variability seem to appear in the common data time range. \cite{liu15} summarized the 5 GHz IDV data of B\(0925+504\) observed prior to 2015, which can be fitted with an ISS model. With the two data points observed in 2015 added, the combined data can also be fit to an updated ISS model. The new model, which may be more reliable because it has been tested against more data, suggests the time scale of \(0.3\pm0.1\) day for B\(0925+504\) is roughly consistent with the annual modulation model of ISS. As for the highly periodic radio-band variations of B\(0925+504\), a tentative interpretation is that the ISS can induce periodic variations when the scintillation pattern forms a periodic distribution of the focus and defocus scintillation patches. The quasi-periodic variations have also been seen at 4.8 and 8.6 GHz in PKS\(1257-326\), an IDV source about half as strong as B\(0925+504\),  by \cite{big03}.

For a source to produce intrinsic radio IDV at 5 GHz with a timescale of 0.5 day, for example, we can estimate the variability brightness temperature \(T_{\rm b,var}\approx1.3\times 10^{18}\) K \citep{wag95} for a flux-density change of 0.1 Jy at 5 GHz during the variability period of 0.5 day for the B\(0925+504\). Based on this \(T_{\rm  b,var}\) and the Doppler effect in total flux variability, \(T_{\rm  b,var}=\delta^{3} T_{\rm int}\) \citep{hov09}, we estimate a Doppler factor \(\delta\ge160\) if we assume an intrinsic brightness temperature \(T_{\rm int}\leq T_{\rm IC}=10^{11.5}\) K. This \(\delta\sim160\) is much greater than that estimated from VLBI data or from total flux variability on longer time scales for blazars \citep[\(\delta<30\),][]{laht99,hov09,sav10,liod17}. The Doppler factor for a test with a viewing angle \(\theta=1^{\circ}\) and an intrinsic jet speed \(v=0.98c\) is \(\delta\sim10\). The intrinsic variations at 5 GHz, with such a short timescale of 0.5 day, could only be possible for a source in the most extreme conditions, e.g. for an extreme case of \(\theta=0.1^{\circ}\) and \(v=0.99992c\), the Doppler factor is \(\delta\sim155\), which is close to the value of 160 obtained above. However, for the B\(0925+504\), as discussed previously an annual cycle surely demonstrates that these radio variations seen in B\(0925+504\) are predominantly caused by ISS.

To summarise, we conducted a pilot campaign of radio- and optical-band IDV observations of five blazars around December 18--21, 2015 with the radio telescope at Effelsberg, and several optical telescopes. After calibration, the light curves from both the 5 GHz radio band and the optical R band were obtained, although the data are not well sampled in time. We tentatively analyse the amplitudes and time scales of the variabilities and any possible periodicity. The blazars vary significantly in the radio (except 3C66A and BL Lac with only marginal variations) and optical bands on intra- and inter-day time scales. No significant correlation between radio- and optical-band variability is found in the five sources, which is most likely due to the radio IDV being dominated by interstellar scintillation whereas the optical variability comes from the source itself. However, the weak correlation between radio- and optical-band variation in some sources, whether real or not, should be investigated again based on observations that yield well-sampled data.

Some gaps appear in the light curves obtained in this pilot campaign, so future radio-optical IDV observations over continuous days with no gaps are required to well identify any correlation between radio- and optical-band variations and the time delay between them. In addition, the optical band is better suited for observing radio-band IDV at both cm and mm wavelengths because the ISS-induced variability is not present at mm wavelengths.

\section*{Acknowledgments}

We gratefully acknowledge the referee for careful reading and valuable comments. This work is supported from the following funds: the 973 Program 2015CB857100, the Key Laboratory of Radio Astronomy, the Chinese Academy of Sciences, the National Natural Science Foundation of China (Grants No. 11273050, No. 11463001 and No. 11503071) and the Guangxi Science Foundation (2014GXNSFAA118024). O.M.K acknowledges financial support from the Shota Rustaveli NSF under contract FR/217554/16; and S. M. Hu acknowledges support from the National Natural Science Foundation of China under Grant No. 11203016 and support from the Young Scholars Program of Shandong University, Weihai. S.Z. acknowledges partial support from  NCN Grant No. 2013/09/B/ST9/00599. Optical monitoring observations of OJ287 and BL Lac collected within the framework of the Blazar Optical Sky Survey (BOSS Project), is conducted at the University of Athens, Greece. We thank TUBITAK for partial support for using the T60 telescope through Project No. 10CT60-76. The radio data are based on observations with the 100-m telescope at the MPIfR (Max-Planck-Institut f\"ur Radioastronomie) at Effelsberg.







\bsp    
\label{lastpage}
\end{document}